\documentclass[%
aps,%
prl,%
superscriptaddress,%
amsmath,amssymb,
reprint,%
]{revtex4-1}

\usepackage{graphicx}
\usepackage{dcolumn}
\usepackage{bm}
\usepackage[colorlinks=true,allcolors=blue]{hyperref}

\newcommand{\RNum}[1]{\uppercase\expandafter{\romannumeral #1\relax}}
\usepackage{upgreek}
\usepackage{times}

\begin{document}
\title{Integrated Server for Measurement-Device-Independent Quantum Key Distribution Network}

\author{Ci-Yu Wang}
\affiliation{State Key Laboratory of Advanced Optical Communication Systems and Networks, School of Physics and Astronomy, Shanghai Jiao Tong University, Shanghai 200240, China}
\affiliation{Synergetic Innovation Center of Quantum Information and Quantum Physics, University of Science and Technology of China, Hefei, Anhui 230026, China}
\author{Jun Gao}
\affiliation{State Key Laboratory of Advanced Optical Communication Systems and Networks, School of Physics and Astronomy, Shanghai Jiao Tong University, Shanghai 200240, China}
\affiliation{Synergetic Innovation Center of Quantum Information and Quantum Physics, University of Science and Technology of China, Hefei, Anhui 230026, China}
\author{Zhi-Qiang Jiao}
\affiliation{State Key Laboratory of Advanced Optical Communication Systems and Networks, School of Physics and Astronomy, Shanghai Jiao Tong University, Shanghai 200240, China}
\affiliation{Synergetic Innovation Center of Quantum Information and Quantum Physics, University of Science and Technology of China, Hefei, Anhui 230026, China}
\author{Lu-Feng Qiao}
\affiliation{State Key Laboratory of Advanced Optical Communication Systems and Networks, School of Physics and Astronomy, Shanghai Jiao Tong University, Shanghai 200240, China}
\affiliation{Synergetic Innovation Center of Quantum Information and Quantum Physics, University of Science and Technology of China, Hefei, Anhui 230026, China}
\author{Ruo-Jing Ren}
\affiliation{State Key Laboratory of Advanced Optical Communication Systems and Networks, School of Physics and Astronomy, Shanghai Jiao Tong University, Shanghai 200240, China}
\affiliation{Synergetic Innovation Center of Quantum Information and Quantum Physics, University of Science and Technology of China, Hefei, Anhui 230026, China}
\author{Zhen Feng}
\affiliation{State Key Laboratory of Advanced Optical Communication Systems and Networks, School of Physics and Astronomy, Shanghai Jiao Tong University, Shanghai 200240, China}
\affiliation{Synergetic Innovation Center of Quantum Information and Quantum Physics, University of Science and Technology of China, Hefei, Anhui 230026, China}
\author{Yuan Chen}
\affiliation{State Key Laboratory of Advanced Optical Communication Systems and Networks, School of Physics and Astronomy, Shanghai Jiao Tong University, Shanghai 200240, China}
\affiliation{Synergetic Innovation Center of Quantum Information and Quantum Physics, University of Science and Technology of China, Hefei, Anhui 230026, China}
\author{Zeng-Quan Yan}
\affiliation{State Key Laboratory of Advanced Optical Communication Systems and Networks, School of Physics and Astronomy, Shanghai Jiao Tong University, Shanghai 200240, China}
\affiliation{Synergetic Innovation Center of Quantum Information and Quantum Physics, University of Science and Technology of China, Hefei, Anhui 230026, China}
\author{Yao Wang}
\affiliation{State Key Laboratory of Advanced Optical Communication Systems and Networks, School of Physics and Astronomy, Shanghai Jiao Tong University, Shanghai 200240, China}
\affiliation{Synergetic Innovation Center of Quantum Information and Quantum Physics, University of Science and Technology of China, Hefei, Anhui 230026, China}
\author{Hao Tang}
\affiliation{State Key Laboratory of Advanced Optical Communication Systems and Networks, School of Physics and Astronomy, Shanghai Jiao Tong University, Shanghai 200240, China}
\affiliation{Synergetic Innovation Center of Quantum Information and Quantum Physics, University of Science and Technology of China, Hefei, Anhui 230026, China}
\author{Xian-Min Jin}
\thanks{xianmin.jin@sjtu.edu.cn}
\affiliation{State Key Laboratory of Advanced Optical Communication Systems and Networks, School of Physics and Astronomy, Shanghai Jiao Tong University, Shanghai 200240, China}
\affiliation{Synergetic Innovation Center of Quantum Information and Quantum Physics, University of Science and Technology of China, Hefei, Anhui 230026, China}
\date{\today}
\begin{abstract}
Quantum key distribution (QKD), harnessing quantum physics and optoelectronics, may promise unconditionally secure information exchange in theory. Recently, theoretical and experimental advances in measurement-device-independent (MDI-) QKD have successfully closed the physical backdoor in detection terminals. However, the issues of scalability, stability, cost and loss prevent QKD systems from widespread application in practice. Here, we propose and experimentally demonstrate a solution to build a star-topology quantum access network with an integrated server. By using femtosecond laser direct writing technique, we construct integrated circuits for all the elements of Bell state analyzer together and are able to integrate several such analyzers on single photonic chip. The measured high-visibility Bell state analysis suggests integrated server a promising platform for the practical application of MDI-QKD network.

\end{abstract} 
\maketitle
Living in a new era of `Internet Of Everything', the ability to build a secure communication network is essential more than ever. Since the emergence of QKD\cite{Bennet1984} -- an indispensable part of quantum communication, many proof-of-principle demonstrations of quantum internet\cite{Elliott2002,Chou2007,Kimble2008,Frohlich2013,Pirandola2016,Maring2017,Humphreys2018} have been done in increasingly complex systems. Experiments have boomed in bulk components implementations\cite{Jin2010,Ji2016,Liao2017,Ren2017}, leaving connecting multiple users an expensive as well as a massive construction project. Integrated quantum device, featuring its cheap, compact, stable and flexible performance, becomes a revolutionary solution to establish quantum communication links.

Much efforts have been put forward on versatile integrated QKD platforms\cite{Tanzilli2012,Politi2008,Metcalf2014}. Chip-based client\cite{Zhang2014,Autebert2016,Melen2016,Sibson2017,Bacco2017,Choe2018} has been developed to investigate the practical possibilities for miniaturization of the quantum communication terminal. Recently, Bunandar {\it et al.}\cite{Bunandar2018} achieved a 43-km-long link with silicon integrated devices, paving the way for one-way long-distance and high-speed chip-based QKD. 

However, these point-to-point schemes have been threatened by many attacks through physical loopholes in detection terminals. MDI-QKD\cite{Lo2012,Da2013,TangZ2014,TangY2014} introduces an independent untrusted third party (Charlie) to announce the measurement results, and the end users (Alice and Bob) only need to encode and send their photons (FIG. 1a). It is immune to all attacks on detection and has a high tolerance of finite basis-dependent flaws. 

In this letter, we propose a solution to build a star-topology\cite{Xu2015,TangY2016,Valivarthi2017,Wang2018} quantum access network with an integrated server. The star topology is found perfectly matching the features of MDI-QKD scheme. All clients only need to build one quantum channel with the server, where an all-optical router can switch and connect any two clients on an integrated Bell state analyzer (FIG. 1b). The keys are not touched in the routing process. Furthermore, this solution is very resource-efficient for multi-user networks. The proposed star network needs $N$ quantum channels to fully connect any two clients, which is far below the requirement of mesh network (FIG. 1c), $C^2_N$ quantum channels, constructed with point-to-point schemes. Meanwhile, this solution needs plenty of Bell state analyzers for simultaneously supporting QKD among many pairs of clients, which, methodologically, can be well solved by constructing integrated circuits for all the elements of Bell state analyzer and assembling an array of such integrated analyzers on single photonic chip in server, as is shown in FIG. 1d.

\begin{figure*}[htb!]
	\centering
	\includegraphics[width=1.85\columnwidth]{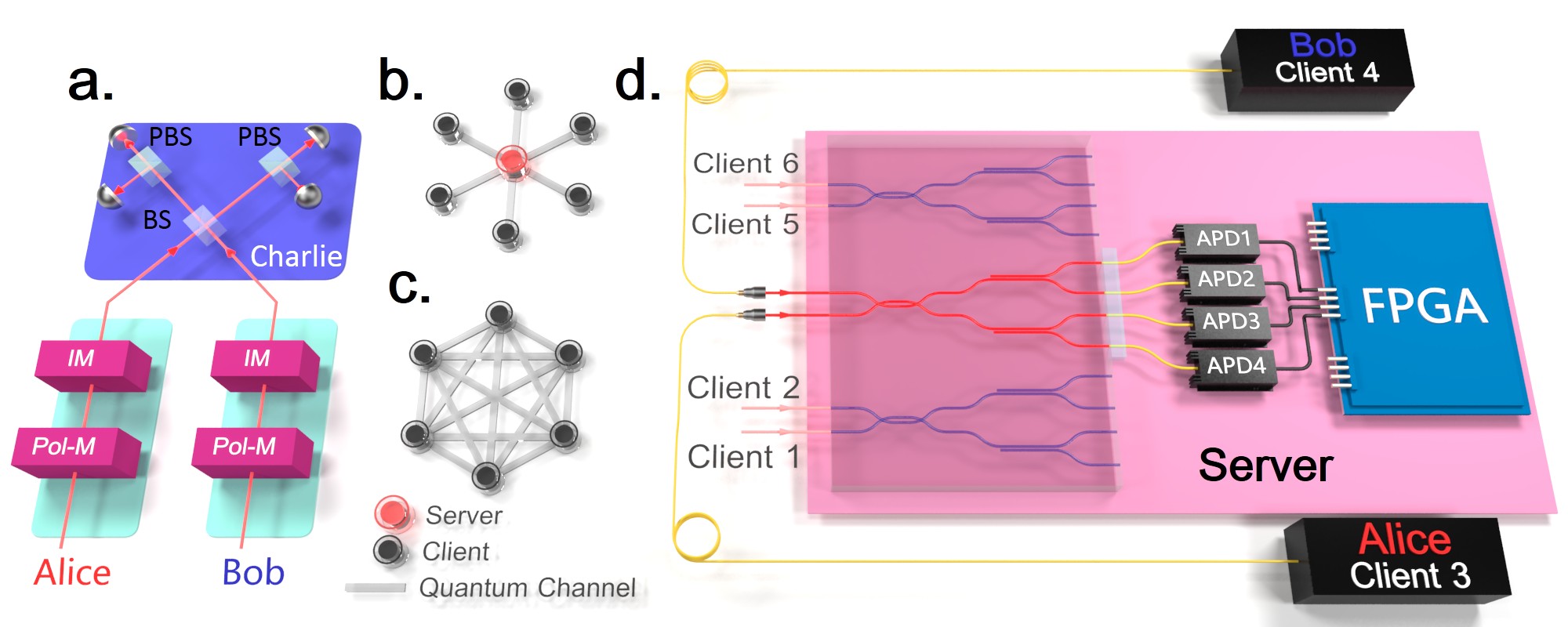}
	\caption{\textbf{Schematic of integrated server for MDI-QKD} \textbf{a.} Overview of MDI-QKD protocol. Two clients (Alice and Bob) encode their photons in BB84 bases by polarization modulator (Pol-M) and send them to an untrusted relay Charlie to conduct Bell state analysis. Intensity modulators (IM) are used to generate decoy states. The Bell state analyzer is composed of one 50/50 beam splitter (BS) and a polarization beam splitter(PBS) on each of its output ports for projection. \textbf{b.} Star-topology network. \textbf{c.} Mesh-topology network. \textbf{d.} Integrated server for proof-of-principle test. Two single photons from Alice and Bob are employed to qualify on-chip Bell state analyzer. Signals from off-chip avalanche photodiode detectors (APD) are processed by a field-programmable gate array (FPGA).}
	\label{Figure 1}
\end{figure*}

\begin{figure}[tb!]
	\centering
	\includegraphics[width=0.98\columnwidth]{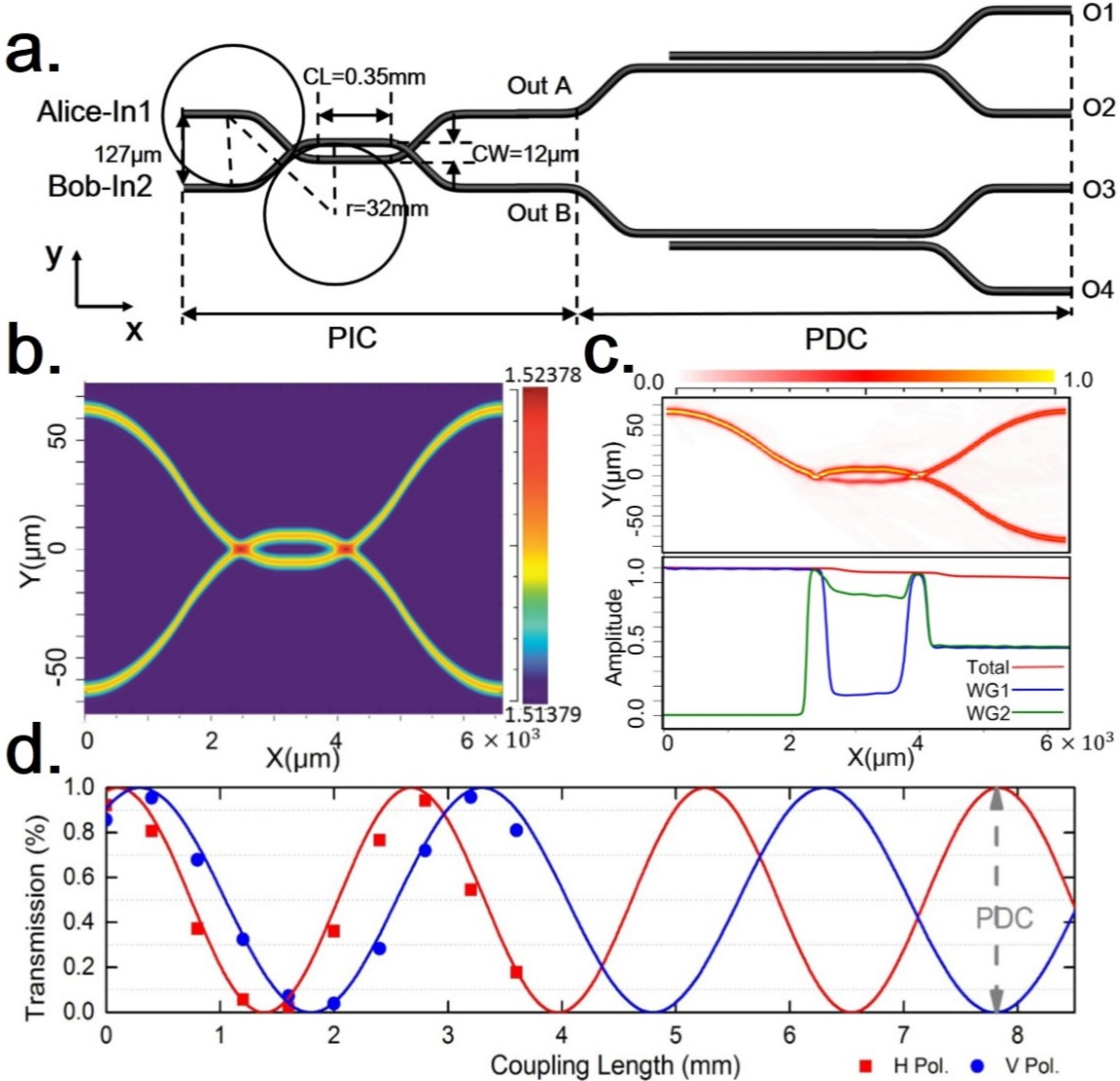}
	\caption{\textbf{Design of integrated elements of on-chip Bell state analyzer.} \textbf{a.} Top view geometry of integrated analyzer consisting of one polarization insensitive coupler (PIC) and two identical polarization directional couplers (PDC). r: radius of fabrication circular arc. CL: coupling length. CW: coupling width. \textbf{b.} Contour map of index simulation of PIC with designed two crossings. \textbf{c.} Power distribution in waveguides with a Gaussian beam injected from In1. \textbf{d.} Measured transmission of couplers by scanning coupling lengths when coupling width is fixed at 8$\upmu$m, for horizontal-(red square) and vertical-(blue circle) polarization input light. Solid lines are the best-fitting curves according to coupling mode theory. Dashed line marks the right coupling length for realizing a PDC.}
	\label{Figure 2}
\end{figure}

We employ femtosecond laser direct writing\cite{Davis1996} to realize all the elements of Bell state analyzer, since the written chip can uniquely be single-mode-fiber compatible for low insertion loss, and can be fabricated in a single-step, mask-free and cost-effective way. By controlling the nonlinear absorption, we are able to tune the mode profile and birefringence of the waveguide for polarization manipulating integrated devices\cite{Sansoni2010,Crespi2011,Fernandes2012,Vest2015,Gao2016,Tang2018}. The waveguides used in this work are fabricated in boro-silicate glass via femtosecond laser direct writing. Waveguides are formed by laser-induced permanent refractive index change, locating at 170 $\upmu$m under the substrate surface by focusing femtosecond laser (wavelength 513nm, repetition rate 1 MHz, pulse energy 210 nJ, writing speed 20 mm/s) through a 50$\times$ objective (numerical aperture 0.55). Profiting from the mode control capacity of femtosecond laser direct writing, the written chip can be linked to a standard V-groove fiber array packaging with facet coupling as low as 1dB.

FIG. 2a shows the detailed design of the Bell state analyzer. The special configuration of polarization insensitve coupler (PIC) is found not very sensitive to fabrication parameters in terms of splitting ratio. The simulation results with commercially available softwares (RSoft BeamPROP and CAD suite) are shown in FIG. 2b and 2c. Polarization insensitiveness is made possible by geometrically controlling interaction length with different coupling width to compensate the birefringence-induced difference of coupling coefficients between horizontal (\textit{H}) and vertical (\textit{V}) polarized light\cite{TPitsios2017,Corrielli2018}. The characterized performance shown in TABLE 1 confirms the high repeatability and uniformity of our PIC. 

According to coupling mode theory, photons transferring from one waveguide to the other follows a sinusoidal law as shown in FIG. 2d. Slight difference in coupling coefficients for \textit{H} and \textit{V} polarized light leads to different oscillation periods, and finally an opposite output behaves as a polarization directional coupler (PDC). Two waveguides are prototyped with a fixed coupling width 8 $\upmu$m, small enough but not overlapped, to ensure that the two propagating modes become coupled during evanescent field overlap.

\begin{figure*}[!htb]
	\centering
	\includegraphics[width=1.5\columnwidth]{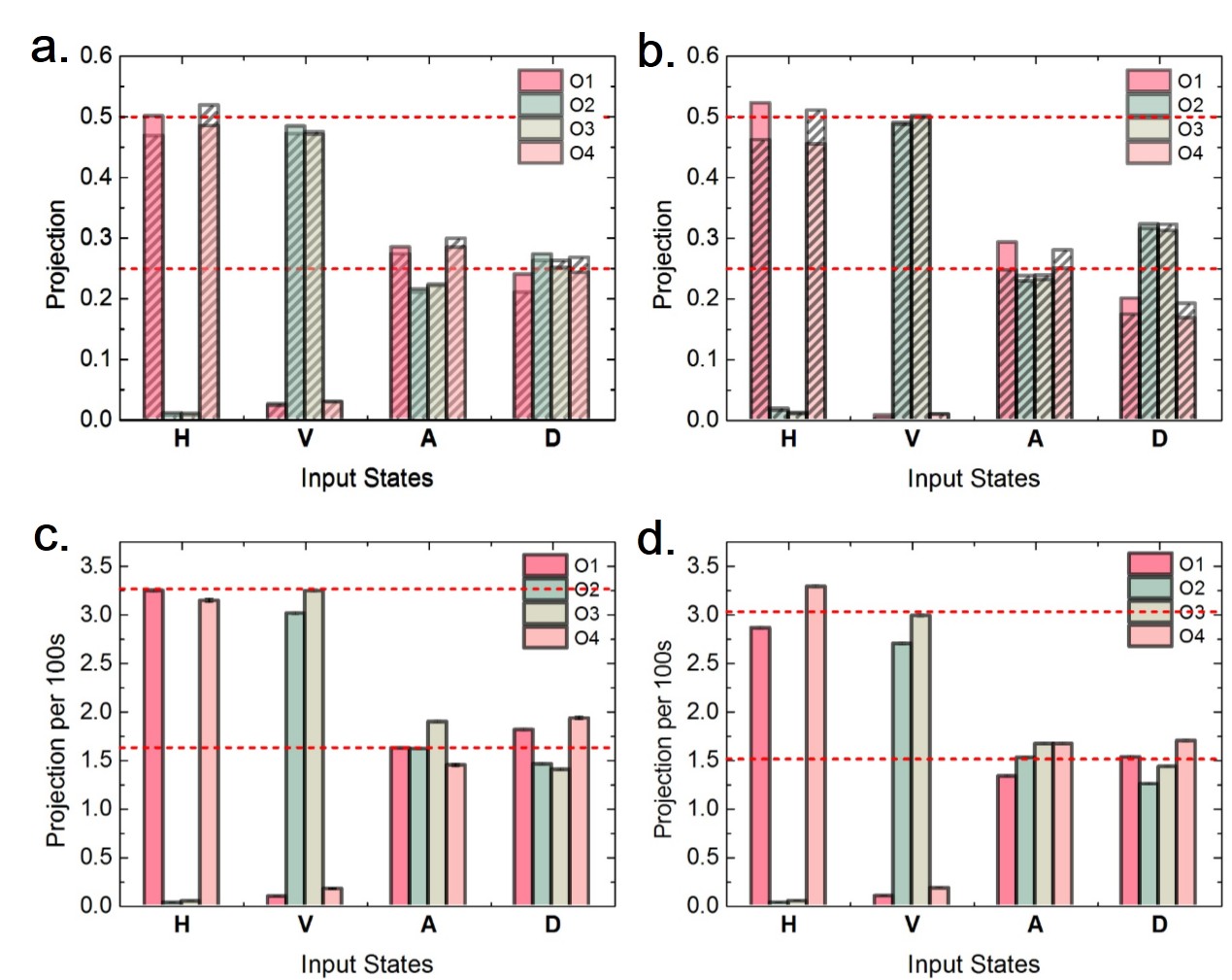}
	\caption{\textbf{Characterized polarization performance of on-chip Bell state analyzer.} \textbf{a.} Measured polarization projection with a probe laser at 780 nm. \textbf{b.} Measured polarization projection with a probe laser at 786 nm. \textbf{c.} and \textbf{d.}  Measured projection coincidence counts per 100 seconds with heralded single photon at 780 nm injected from In1 and In2. }
	\label{Figure 3}
\end{figure*}

\begin{table}[!t]
	\centering
	\caption{\textbf{50/50 PIC performance.} Output splitting ratios under different input conditions.}
	\label{tab1}
	\begin{tabular}{p{2.1cm}<{\centering} p{1.5cm}<{\centering} p{1.5cm}<{\centering}  p{1.5cm}<{\centering} p{1.5cm}<{\centering}}
		\hline\noalign{\vskip 0.6mm}
		\hline\noalign{\vskip 0.14cm}
		Input States &  In1-OutA  & In1-OutB & In2-OutA & In2-OutB\\[0.07cm]
		\hline\noalign{\vskip 0.12cm}
		H			& 50.4  &49.6  &48.6  &51.4       \\[0.06cm]
		V			& 48.2  &51.8  &50.7  &49.3       \\[0.06cm]
		A			& 49.4  &50.6  &50.4  &49.6       \\[0.06cm]
		D			& 48.7  &51.3  &50.1  &49.9       \\[0.06cm]
		\hline\noalign{\vskip 0.6mm}
		\hline\noalign{\vskip 0.6mm}
	\end{tabular}
	\par
\end{table}

\begin{figure}[htb!]
	\centering
	\includegraphics[width=1\columnwidth]{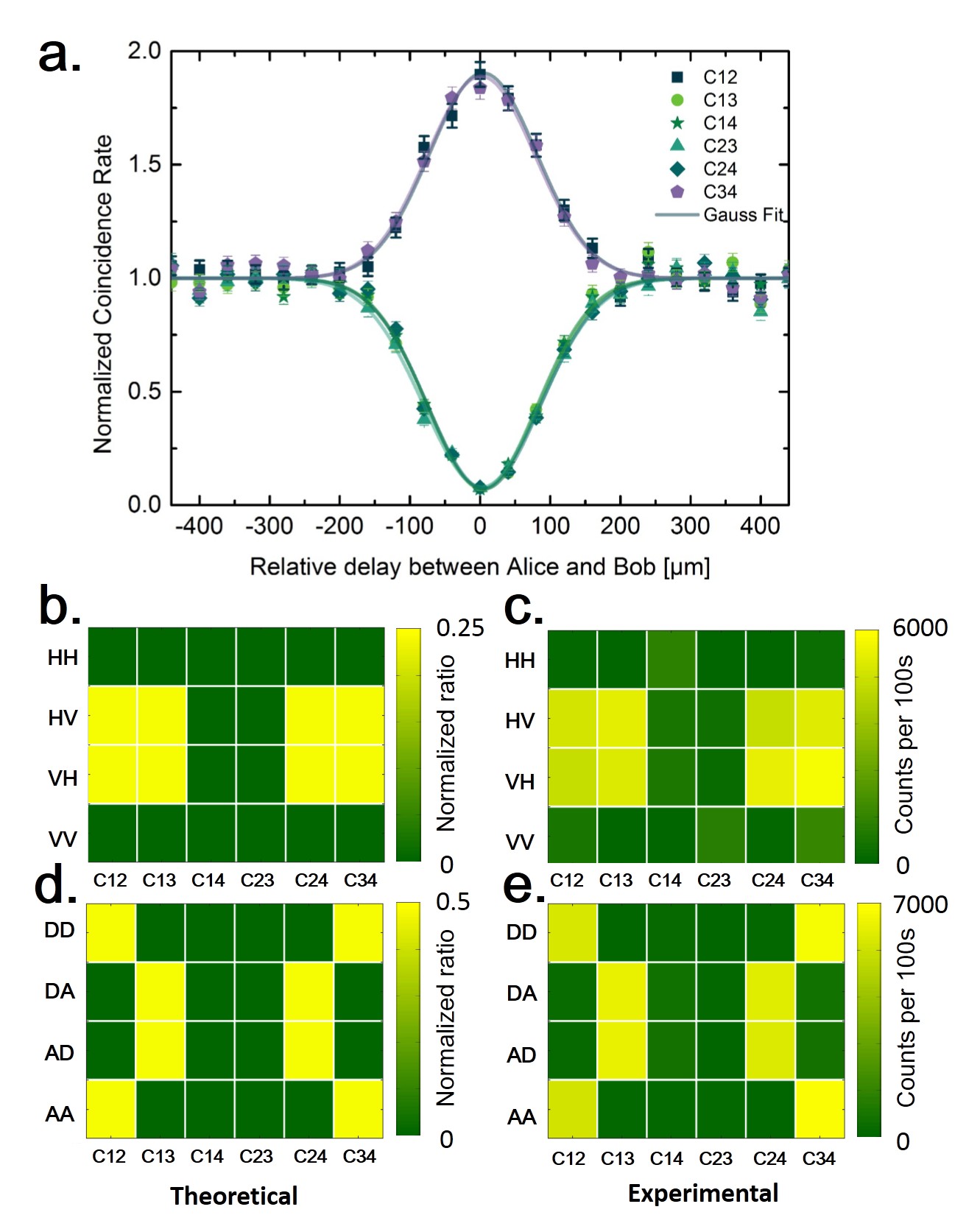}
	\caption{\textbf{Experimental results of Bell state analysis with integrated server.} \textbf{a.} Hong-Ou-Mandel Interference for all the coincidence counts. An average visibility over 90$\%$ is reached. \textbf{b.-e.} Relative conditional probability distribution with different input states in rectilinear and diagonal bases. \textbf{b.} and \textbf{d.} show the theoretical distributions and \textbf{c.} and \textbf{e.} show the experimental coincidence distributions.}
	\label{Figure 4}
\end{figure}

The main task of MDI-QKD is to conduct Bell state analysis, thus we first use a probe laser to characterize our integrated analyzer. The results obtained by injecting photons from In1 and In2 are presented with hatched and colored histograms respectively, and are piled together to clearly reveal their uniformity for different inputs and for different wavelengths of light at 780 nm (FIG. 3a) and 786 nm (FIG. 3b). Our probe laser can only be tuned in a limited range, and however, the good performance at two different wavelengths implies the ability to fabricate our integrated analyzer in telecom band through similar parameter optimization method.

To test the performance of polarization projection on two PDCs, we prepare four typical input states for BB84 protocol bases, polarized at 0$^\circ$ (\textit{H}), 90$^\circ$ (\textit{V}), 45$^\circ$ (\textit{D}) and 135$^\circ$ (\textit{A}). The normalized projection results agree with the expected relative ratio for \textit{H/V} and \textit{A/D} bases (red dashed lines). More importantly, the average polarization extinction ratio for \textit{H/V} bases is about 20:1, which can well identify $|\psi^+\rangle$ out of $|\phi^\pm\rangle$ for Bell state analysis.

However, the response and noise level of detection system for laser and single photons are different. Thus, we repeat the experiment with input states of heralded single photons. We generate photon pairs via Type-\RNum{2} spontaneous parametric down-conversion by pumping a beta-barium borate (BBO) crystal with a Ti:sapphire laser. Heralding efficiency of conditionally detecting one single photon by the other one is over 25$\%$. Apparently, the detection and coupling efficiencies for different outputs can not be identical in practice, which may modify the projection results. As we can see, the projection coincidence counts per 100 seconds shown in FIG. 3c and 3d slightly deviate from the results shown in FIG. 3a, but, are all around the expected relative ratio (red dashed lines). Interestingly, the average polarization extinction ratio for \textit{H/V} bases is as high as 59:1, which of course reflects the real performance of the device at single-photon level.

In order to qualify the integrated server suitable for MDI-QKD, we couple the pair of photons into one of on-chip Bell state analyzers, as is shown in FIG. 1d, both prepared in \textit{D}, and scan their relative delay to find the zone of Hong-Ou-Mandel Interference\cite{Ou2007}. Photons from Alice and Bob are encoded independently in BB84 protocol bases by rotating half wave plates. Since our two PDCs can project incoming states \textit{H} and \textit{V} at the output ports 1, 4 and 2, 3 respectively, coincidence measurements could deterministically discriminate the states $|\psi^+\rangle$ and $|\psi^-\rangle$ by two-fold coincidences C12, C34 and C13, C24 via bunching or anti-bunching effect\cite{Silva2013}. Meanwhile, two-fold coincidences C14 and C23 can not distinguish the states $|\phi^+\rangle$ and $|\phi^-\rangle$, and therefore are not used as successful events in MDI-QKD protocols\cite{Calsamiglia2001}. The Visibilities for all combinations are $89.8\%, 92.4\%, 93.0\%, 92.3\%, 92.2\%$ and $83.7\%$ for C12, C13, C14, C23, C24 and C34 (FIG. 4a).

Two-photon data are collected for eight linearly independent input conditions FIG. 4b and 4c (\textit{HH, HV, VH, VV}), and FIG. 4d and 4e (\textit{DD, DA, AD, AA}), with half wave plates and polarization compensation. The gain values at each rectilinear ($Q_{i,j}^{r}$) and diagonal ($Q_{i,j}^{d}$) basis can be caculated from the coincidence measurements by\cite{Da2013}:
\begin{equation}Q_{i,j}^{r}=\frac{1}{4}(C_{Sum}^{HH}+C_{Sum}^{VV}+C_{Sum}^{HV}+C_{Sum}^{VH})\end{equation}
\begin{equation}Q_{i,j}^{d}=\frac{1}{4}(C_{Sum}^{++}+C_{Sum}^{--}+C_{Sum}^{+-}+C_{Sum}^{-+})\end{equation}
\textit{i} and \textit{j} represent the different average coincidence counts used by Alice and Bob while we adopt the successful events by:
\begin{equation}C_{Sum}^{AB}=C_{12}^{AB}+C_{34}^{AB}+C_{13}^{AB}+C_{24}^{AB}\end{equation}
We get the final gain values: $Q_{i,j}^{r}=1.37\times10^{-6}$ bits/pulse, $Q_{i,j}^{d}=1.75\times10^{-6}$ bits/pulse.
We estimate the quantum bit error rate (QBER) by:
\begin{equation}E_{i,j}^{r}=\frac{C_{Sum}^{HH}+C_{Sum}^{VV}}{C_{Sum}^{HH}+C_{Sum}^{VV}+C_{Sum}^{HV}+C_{Sum}^{VH}}\end{equation}

\begin{equation}
	\begin{aligned}
	E_{i,j}^{d}=&\frac{C_{13}^{++}+C_{24}^{++}+C_{13}^{--}+C_{24}^{--}}{C_{Sum}^{++}+C_{Sum}^{--}+C_{Sum}^{+-}+C_{Sum}^{-+}}\\
	+&\frac{C_{12}^{+-}+C_{34}^{+-}+C_{12}^{-+}+C_{34}^{-+}}{C_{Sum}^{++}+C_{Sum}^{--}+C_{Sum}^{+-}+C_{Sum}^{-+}}
	\end{aligned}
\end{equation}

From our experimental data, we estimate $E_{i,j}^{r}=0.058, E_{i,j}^{d}=0.081$. The nonideal QBER can be attributed to multi-photon events of the testing photon source, residual birefringence in PIC and polarization-dependent loss. The last two points are being improved to meet the stringent requirement for low QBER and therefore for high final bit rate.

In summary, we propose and experimentally demonstrate an integrated server for MDI-QKD network and its compatibility to a star-topology quantum access network. Every client sends their encoded photons to server through their sole channel. Server can link any two clients simply by switching their photons to one of Bell state analyzers on a photonic chip. All the combination between two clients is the same as a standard MDI-QKD and therefore is unconditionally secure. The number of required channels is much less than mesh-topology with point-to-point schemes for realizing fully connected MDI-QKD networks, which is particularly significant when the networks go to large scales.\\

\textbf{Acknowledgements:} The authors thank Jian-Wei Pan, Hang Li, Xiao-Ling Pang, and Jian-Peng Dou for helpful discussions. This work was supported by National Key R\&D Program of China (2017YFA0303700); National Natural Science Foundation of China (NSFC) (11374211, 61734005, 11690033); Shanghai Municipal Education Commission (SMEC)(16SG09, 2017-01-07-00-02-E00049); Science and Technology Commission of Shanghai Municipality (STCSM) (15QA1402200, 16JC1400405). X.-M.J. acknowledges support from the National Young 1000 Talents Plan.

%

\end{document}